\documentclass[12pt]{article}
\usepackage{latexsym}
\usepackage{epsfig}

\usepackage{graphicx}
\usepackage{epstopdf}

\usepackage{epsfig,amssymb,amsmath,amscd}

\newcommand{\bq}{\begin{equation}}
\newcommand{\eq}{\end{equation}}
\newcommand{\bqa}{\begin{eqnarray}}
\newcommand{\eqa}{\end{eqnarray}}
\newcommand{\ben}{\begin{enumerate}}
\newcommand{\een}{\end{enumerate}}
\newcommand{\bc}{\begin{center}}
\newcommand{\ec}{\end{center}}
\newcommand{\bqb}{\begin{eqnarray*}}
\newcommand{\eqb}{\end{eqnarray*}}

\def\pr#1#2#3{Phys. Rev. ${\bf{#1}}$, #2 (#3)}

\def\pl#1#2#3{Phys. Lett. ${\bf{#1}}$, #2 (#3)}

\def\np#1#2#3{Nucl. Phys. ${\bf{#1}}$, #2 (#3)}

\def\zp#1#2#3{Z. f. Phys. ${\bf{#1}}$, #2 (#3)}

\def\jmp#1#2#3{J. Mod. Phys. ${\bf{#1}}$, #2 (#3)}

\begin{document}
\pagenumbering{arabic}
\thispagestyle{empty}
\def\thefootnote{\fnsymbol{footnote}}
\setcounter{footnote}{1}

\begin{flushright}
April 25, 2017\\
 \end{flushright}

\begin{center}

{\Large {\bf  CSM analyses of $e^+e^- \to t\bar t H, t\bar t Z, t\bar b W$}}.\\
 \vspace{1cm}
{\large F.M. Renard}\\
\vspace{0.2cm}
Laboratoire Univers et Particules de Montpellier,
UMR 5299\\
Universit\'{e} Montpellier II, Place Eug\`{e}ne Bataillon CC072\\
 F-34095 Montpellier Cedex 5, France.\\
\end{center}

\vspace*{1.cm}
\begin{center}
{\bf Abstract}
\end{center}

We study the modifications of the amplitudes and cross sections of several processes, especially 
$e^+e^- \to t\bar t H, t\bar t Z, t\bar b W$, generated by Higgs boson and top quark compositeness, in particular within the CSM concept. We illustrate the observable differences that may appear between various,
CSM conserving or CSM violating, compositeness possibilities.

\vspace{0.5cm}

\def\thefootnote{\arabic{footnote}}
\setcounter{footnote}{0}
\clearpage

\section{INTRODUCTION}

The concept of Composite Standard Model (CSM) is motivated by the fact that
in spite of its great success the Standard Model (SM) presents some definite lacks
which had led to several types of BSM studies. Among them one finds the possibility
of compositeness (substructures or additional sets of states), see for example 
ref.\cite{comp, Hcomp2, Hcomp3, partialcomp, Hcomp4},
which would especially affect the Higgs boson and the top quark.\\
The basic CSM hypothesis is that the structure and the main properties of the 
SM are preserved, at least at low energy,
in spite of the fact that the Higgs boson and the top quark would be composite particles.\\
Our aim is not to analyze a specific compositeness model,
for example one of those mentioned in the above references, but to see 
how various observables would be sensitive to such types of effects, for example to
the presence of form factors which are close to one at low energy.\\
The CSM analysis could proceed in the following way.
One starts from the $e^+e^-\to ZH$ process and looks for the existence of 
a $ZZH$ form factor, see \cite{WLZL}.\\
Similarly one looks for the presence of left and right top quark form factors 
in  $e^+e^-\to t\bar t$.\\
The next step would be to check the SU(2)*U(1) structure of these form factors, which means
on the one hand to check
if the presence of a Higgs form factor can be generalized to a set of $G^{\pm,0}$ form factors 
transmitted to $W^{\pm}_L$, $Z_L$,\
to be for example confirmed by an analysis of $e^+e^-\to W^+W^-$,
and on another hand to check if they are in some way related to the top quark ones; this could
be natural if the top quark and the Higgs boson have the same subconstituents. In fact
such relations may be imposed by the energy behaviour of the $ZH$  one loop production amplitudes
in $gg$ and $\gamma\gamma$ collisions. 
In \cite{CSMgg,CSMgamgam} we have remarked that the $gg\to ZH$ and $\gamma\gamma\to ZH$
processes are particularly sensitive to the presence of form factors because they could 
destroy a peculiar SM cancellation between diagrams involving Higgs boson and top quark
couplings. But we have also shown that this cancellation can be preserved provided 
a special relation between form factors is satisfied. We considered this
relation as a specific CSM property.
The same constraints can be inferred by looking directly at the high energy behaviour 
of the $t\bar t\to Z_LH$ amplitudes.\\
This procedure can be generalized to $t\bar t$ production amplitudes in ZZ 
and WW collisions, especially with longitudinal Z and W which could be composite. 
But as mentioned in \cite{trcomp}, in such processes where the resulting SM amplitudes 
appear to be proportional to the top mass $m_t$ there is an alternative possibility,
which may be also considered as consistent with the CSM concept, 
which consists in introducing an effective scale dependent
mass $m_t(s)$ which would ensure a good high energy behaviour.\\

In the present paper we want to extend such CSM analyses to further processes involving
top quarks and/or Higgs and Goldstone bosons (i.e. $W_L,Z_L$). We will affect form
factors to their couplings and look at the behaviours of the amplitudes when specific
CSM constraints are imposed and/or when the $m_t(s)$ dependence is applied. 
Detailed applications will be given for $t\bar t H$, $t\bar t Z$, $t\bar b W$ 
production in $e^+e^-$ collisions.\\

Contents:  In Section 2 we recall the definitions of the various Higgs boson 
and top quark form factors and the possible CSM constraints. The precise affectations
of the form factors are listed in Section 3 following the structure of the
SM diagrams describing the $e^+e^- \to t\bar t H, t\bar t Z, t\bar b W$ amplitudes.
Illustrations of their consequences for the ratios of modified cross sections 
over the SM ones are given in Section 4 and the conclusion in Section 5.\\

\section{BASIC HIGGS AND TOP QUARK FORM FACTORS AND CSM CONSTRAINTS}

Higgs boson compositeness should reveal itself by the presence of 
form factors in specific $H$ couplings.
What we mean by "form factor" is a modification of the point-like coupling.
It may have any kind of shape, an increase due to new contributions (resonances,
excited states, ...), or a simple decrease due to the an extended spatial structure.
There are no $\gamma$ nor $Z$ coupling to a pure Higgs boson $H$.
The simplest place concerning a pure $H$ form factor would be the $HHH$ coupling,
not for looking for anomalous components, but for an s-dependence when
one $H$ line is off-shell; this is difficult to measure, see \cite{mumuH}.\\

However if compositeness preserves the whole SM Higgs doublet structure then
form factors may affect the following 
$ZG^0H$, $W^{\pm}G^{\pm}G^0$, $W^{\pm}G^{\pm}H$, $HG^{\pm,0}G^{\pm,0}$, 
$ZG^{\pm}G^{\pm}$, $\gamma G^{\pm}G^{\pm}$ 
couplings and, if the equivalence is preserved as assumed by CSM, the corresponding ones
when  $G^{\pm,0}$ are replaced by $W^{\pm}_L$, $Z_L$.\\
The simplest case case is probably the observation of a $ZZ_LH$ form factor
equivalent to a $ZG^0H$ one in the $e^+e^-\to ZH$ process, see \cite{WLZL}.\\
A test of the preservation of the Goldstone equivalence can then be done
in $e^+e^-\to W^+W^-$. A well-known SM feature is the cancellation of the $t$ channel
neutrino exchange contribution with the $\gamma,Z$ $s$ channel exchange one
to the $W^+_LW^-_L$ amplitude at high energy. A crude form factor effect in
the  $\gamma,Z-W^+W^-$ couplings destroys this cancellation and leads to
an inacceptable increase of the $e^+e^-\to W^+_LW^-_L$ amplitude.
However if the Goldstone equivalence is preserved in some (CSM) fashion, 
the high energy $e^+e^-\to W^+_LW^-_L$ amplitude remains equal to the
$e^+e^-\to G^+_LG^-_L$ one (only due to $\gamma,Z$ $s$ channel exchange),
so that even if a form factor affects the  $\gamma,Z-G^+G^-$ coupling
this amplitude can behave correctly.\\

Independently, through the $e^+e^-\to t\bar t$ process, one can obviously 
detect the presence of 
$\gamma t_{L,R}t_{L,R}$ and $Zt_{L,R}t_{L,R}$ form factors with the
options of both  $t_{L,R}$ or of only $t_{R}$ compositeness, see \cite{trcomp}.\\

The next step at which CSM constraints appear are the $gg\to ZH$ and  
$\gamma\gamma\to ZH$ processes occuring via 1 loop triangle and box quark loops,
\cite{CSMgg,CSMgamgam}. One can equivalently observe them in the high energy 
behaviour of the $t\bar t\to Z_LH$ amplitudes. 
In order to not generate a strong enhancement
one should preserve a typical SM cancellation of the contributions 
of the top (t- and u- channel) exchange and of the $G^0$ s-channel exchange.\\
Introducing five arbitrary effective form factors chosen as
$F_{G^0Z_LH}(s)=F_{ZZ_LH}(s)$, $F_{Htt}(s)$, $F_{Gtt}(s)$, 
$F_{tR}(s)$, $F_{tL}(s)$ this preservation occurs provided the following 
CSM constraint is satisfied:
\bq
F_{G^0Z_LH}(s)F_{Gtt}(s)(g^Z_{tR}-g^Z_{tL})=
F_{Htt}(s)(g^Z_{tR}F_{tR}(s)-g^Z_{tL}F_{tL}(s))~~\label{CSMconsZH}
\eq

In a similar fashion one can then consider first the $t\bar t\to Z_LZ_L$
process giving the constraint
\bq
-{1\over2}g_{HZZ}g_{Htt}F_{HZZ}(s)F_{Htt}(s)=
m_{t}((g^Z_{tR}F_{tR}(s)-g^Z_{tL}F_{tL}(s))^2~~\label{CSMconsFFZ}
\eq
\noindent
and the $t\bar t\to W_LW_L$ processes  requiring
\bq
F_{HWW}(s)F_{Htt}(s)=
F^2_{tL}(s)=F_{VWW}(s)F_{tL}(s)~~\label{CSMconsFFW}
\eq
In this second case in order to recover the SM structure one needs
to require that the $\gamma tt$ and $Ztt$ form factors are similar as well
as the $F_{\gamma WW}(s)$ and $F_{ZWW}(s)$ form factors. Finally the two
above constraints require that all the involved form factors have a common 
$F(s)$ shape.\\
However one should note that each of these contributions to the amplitude
appears kinematically proportional to the top mass such that an alternative 
way to reduce their growing behaviour when the cancellation does 
not directly occur would be to replace $m_t$ by an effective (compositeness) 
quantity $m_t(s)$, see \cite{trcomp}.\\

One can pursue this type of study with other processes involving Higgs bosons and/or
longitudinal gauge bosons and/or top quarks.\\
A first set corresponds to the famous WW, WZ, ZZ scattering occuring through
gauge and Higgs bosons exchanges. Introducing form factors for $H$, $W_L$ and $Z_L$
couplings, the cancellation would also require that they have a common $F(s)$ shape.\\
 
Another set which consists of  $e^+e^-\to WWH, ZZH$ processes should give the same 
type of CSM constraint.\\
Finally a richer set is provided by  $t\bar t H$, $t\bar t Z$ and $t\bar b W$
processes in $e^+e^-$ or in hadronic collisions because they directly involve
both the Higgs boson and the top quark sectors.\\

Careful analyses of each of these processes should determine the detailed properties
of the $H$ and top quark compositeness and in particular to see if the
modifications of the SM predictions correspond to
CSM conservation or to CSM violation in each of these sectors.\\
With this aim in the next sections we present a possible analysis of these processes 
in $e^+e^-$ collisions which have been considered since a long time \cite{ttH,ttZ}. 
We will show that the  $t\bar b W$ case, maybe more difficult to observe,
would be the more sensitive to such effects.\\

\section{TREATMENT OF THE $e^+e^-\to t\bar t H, t\bar t Z, t\bar b W^-$ PROCESSES}

\subsection{$e^+e^-\to t\bar t H$}

This proceeds through two types of diagrams, see the first set of Fig.1:\\
--- (a),(b) for  s-channel $\gamma,Z$ exchange creating a $t\bar t$ pair followed
by $H$ emission  from $t$ or from $\bar t$;\\
--- (c) for s-channel  $Z$ exchange creating $HZ$ followed by $Z\to t\bar t$;
the second intermediate $Z$ involves the $G^0$ contribution.\\ 

The couplings that can be affected by Higgs and top compositeness are 
$\gamma tt$, $Ztt$, $Htt$ and $ZZH$ (involving $ZG^0H$).\\
A specific form factor may be attributed to each of them as discussed
in the previous section, with or without CSM constraint.\\
In the illustrations we will introduce
$F_{tR}(s)$, $F_{tL}(s)$ (the same for $\gamma$ and $Z$), $F_{Htt}(s)$
and $F_{ZG^0H}(s)$. \\

\subsection{$e^+e^-\to t\bar t Z$}

The second set of diagrams of Fig.1 concerns this process and is composed of:\\ 
--- (a),(b) for  s-channel $\gamma,Z$  exchange creating a $t\bar t$ pair followed
by $Z$ emission  from $t$ or from $\bar t$;\\
--- (c) s-channel  $Z$ exchange creating $HZ$ followed by $H\to t\bar t$;\\
--- (d),(e) for $Z$ emission by $e^+$ or $e^-$  line and $\gamma,Z$ from the other line
creating a $t\bar t$ pair.\\

The couplings to discuss are again the $\gamma tt$, $Ztt$, $Htt$ and $ZZH$,
with the form factors $F_{tR}(s)$, $F_{tL}(s)$, $F_{Htt}(s)$
and $F_{ZG^0H}(s)$, but they appear in rather different combinations. \\

\subsection{$e^+e^-\to t\bar b W^-$}

The third set of diagrams of Fig.1 are:\\
--- (a),(b) for s-channel $\gamma,Z$ exchange creating a $b\bar b$ 
or a $t\bar t$ pair followed
by $W^-$ emission from $b$ or from $\bar t$;\\
--- (c)  for s-channel  $\gamma,Z$ exchange creating a $W^+W^-$ pair
followed by $W^+\to t\bar b$,\\
--- (d) for $W^-$ emission by $e^-$  line and  $W^+\to t\bar b$ from the  $e^+$  line.\\

This process is particularly interesting (as in other previous cases involving
$W^{\pm}_L$) due to the occurence of a special cancellation of the four (a-d) 
contributions (each of them growing like $p/m$) which can be perturbed by form factors. 
The $\gamma bb$, $Zbb$ appearing in (a) should not be affected by form factors
if the bottom quark is not a composite particle. 
The pure left $Wtb$ coupling may have a specific $F_{tL}(s)$ form factor.
The $\gamma WW$, $ZWW$ couplings may have form factors when $W_L$ or $Z_L$
are concerned, i.e. $F_{\gamma W_LW_L}(s)$ and $F_{ZW_LW_L}(s)$ equivalent to
$F_{\gamma GG}(s)$ and $F_{ZGG}(s)$, 
if the $G^{\pm}\simeq W^{\pm}_L$ equivalence is preserved by CSM.\\

\subsection{CSM-conserving versus CSM-violating form factors}

In the next section we will make illustrations showing the effects
obtained in the CSM-conserving case (those which satisfy the constraints given
in Section 2) and in the CSM-violating case (with arbitrary form factors).\\

For simplicity we will use as a simple "test" expression  
\bq
F(s)={(m_Z+m_H)^2+M^2\over s+M^2}~~\label{FF}
\eq
\noindent
with the new physics scale $M$ taken for example with the value of 5 TeV.\\
We insist that the aim of this choice is only to illustrate the sensitivity
of the observables to compositeness effects, but as mentioned in Section 2,
it has no quantitative meaning. The compositeness structure may be much richer 
and require a more involved effective adequate form factor.\\
 
Two CSM conserving cases will be illustrated with the simple choices:\\ 

--- (a) denoted CSMtLR, refering to both $t_{L}$ and $t_{R}$ compositeness, with
$F_{tR}(s)=F_{tL}(s)\equiv F(s)$, this form factor being also present in the
$Wtb$  coupling,
and $F_{G^0Z_LH}(s)=F_{VG^+G^-}(s)=F_{Gtt}(s)=F_{Htt}(s)\equiv F(s)$ satisfying (\ref{CSMconsZH}),
the top mass being fixed at its on-shell value.\\
 
--- (b) denoted CSMtR, for pure $t_R$ compositeness, $F_{tR}(s)\equiv F(s)$ while $F_{tL}(s)=1$,
$F_{VG^+G^-}(s)=1$ and $m_t(s)=m_tF(s)$ controlling all Higgs sector couplings and the top kinematical
contributions.\\

and two CSM violating cases:\\

--- (c) denoted CSMvt, where $F_{tR}(s)$ with $M=10$ TeV differs from $F_{tL}(s)$ with $M=15$ TeV
(also present with the $Wtb$ coupling)
and 
$F_{G^0Z_LH}(s)=F_{VW^+_LW^-_L}(s)=F_{Htt}(s)=F_{Gtt}(s)\equiv F(s)$ with $M=5$ TeV,
the top mass being fixed at its on-shell value.\\

--- (d) denoted CSMvH, with no top form factor and only one  $F(s)$ form factor affecting the 
$G^0Z_LH$, $ZZ_LH$, $VW^+_LW^-_L$ vertices, the top mass being also fixed at its on-shell value.\\

\section{ILLUSTRATIONS}

We now illustrate the changes that the form factors generate on the production cross sections by plotting the ratios of the modified cross section over the SM one in the central domain of the phase space.\\
A priori a form factor like the one in (\ref{FF}) should generate a strong decrease of the couplings and consequently of the amplitudes at high energy. We will see that this indeed happens in the CSM cases which respect the SM structures and do not create anomalous contributions destroying possible specific SM combinations or cancellations. On the opposite CSM violating cases lead to weaker decreases and even to strong enhancements in some cases.\\

\underline{$e^+e^-\to t\bar t H$}\\

As one can see in Fig.2 the form factor reduction effect mentioned at the end of the 
previous section is totally respected in case (a) which correspond to a full CSM case. 
The reduction is somewhat smaller in case (b) with only $t_R$ compositeness. 
It is much smaller in the CSM violating cases (c) and (d).\\

\underline{$e^+e^-\to t\bar t Z$}\\

We treat separately the 3 options corresponding to $Z_L$, $Z_T$, and unpolarized $Z$  
production in Fig.3a,b,c.\\
For $Z_L\simeq G^0$, as expected, the ratio  behaves rather similarly to the 
$t\bar t H$ one, but $Z_T$ is much less affected, especially not by CSMvH as it does not involve the Higgs sector. As $Z_T$ production in SM represents 75 percent of the total at high energy, the unpolarized $Z$ shapes looks like $Z_L$ and $H$, although the global size of the effects is weaker.\\

\underline{$e^+e^-\to t\bar b W^-$}\\

We also separate the 3 options for  $W_L$, $W_T$ and unpolarized $W$ production in Fig.4a,b,c.\\
These options are particularly interesting because of the occurence of specific 
SM combinations of the different diagrams and especially the cancellations of their contributions
to the $W^-_L$ amplitudes which grow individually like $p/m$.
These cancellations are perturbed by the presence of specific form factors for the 3-boson couplings, 
especially in the CSM violation cases CSMvt or CSMvH and this leads to a strong increase 
of the total amplitude, as it can be seen in Fig.4a.\\
On another hand, CSM conserving cases are discussed by using the goldstone equivalence, replacing
$W^-_L$ by $G^-$ and describing these longitudinal amplitudes by the only 3 first diagrams 
(a-c) as the $G^-$ emission in (d) vanishes. In these cases form factors now reduce the size 
of each term without cancellation problem.\\
For similar reasons the $W_T$ amplitudes are only moderatly affected by the form factors as no
cancellation effect occurs, see Fig.4b.\\
Consequently the unpolarized $W$ shapes in Fig.4c are essentially dominated by the strong $W_L$ 
modifications as one can see in Fig.4c.\\

\section{CONCLUSIONS}

In spite of its success, the SM with its large number of free parameters
involving very different scales may suggest the existence of some kind of (simple?) substructure.
The heavier particles like the top quark and the Higgs boson could be the best
places for observing compositeness signals.\\
One another hand it would be adequate that this compositeness preserves the SM structure
at low energy (and does not generate anomalous couplings,....) and that new features 
only progressively appear when the energy increases.  This is the basis of what we
called CSM and of the analyses that we developed with form factors. \\ 

In this paper we have shown that the 
$e^+e^- \to t\bar t H, t\bar t Z,  t\bar b W$ processes are particularly sensitive to compositeness in the top and Higgs boson sectors (i.e. $t_{L}$, $t_R$, $H$, $W_L$, $Z_L$) and should allow to differentiate between various possibilities especially those which preserve the SM structures, and respect the CSM constraints,
and those which violate them.\\ 
We have illustrated these possibilities in Fig.2-4 by using test form factors but these results
have no real quantitative meaning. These effects would be smaller if the new physics scale $M$ is higher. 
They could be locally larger if resonances or excited states contribute in addition to such 
simple form factors.\\ 
But in any case we have shown that indeed this set of 3 processes can be very rich in informations about $t_{L,R}$ and $H$ compositeness. Joined to those that can be obtained from $ZH$, $W^+W^-$ and $t\bar t$ production processes they should allow to select the type of compositeness responsible for the 
modifications and to suggest which new structures (CSM conserving or violating) are involved.
Among the CSM possibilities one may for example identify an effective, scale dependent, top mass $m_t(s)$,
a possibility whose further consequences should be explored. \\

Additional studies could be done, first through $t\bar t H, t\bar t Z,  t\bar b W$ production 
in hadronic \cite{Richard} or in photon-photon collisions, see \cite{trcomp}, where the
top and Higgs couplings appear in combinations differing from those considered in the
present paper.
Very precise studies of processes involving other quarks or leptons could independently reveal a common
(possible partial) compositeness origin.\\

\clearpage

\begin{figure}[p]
\vspace{-6cm}
\[
\epsfig{file=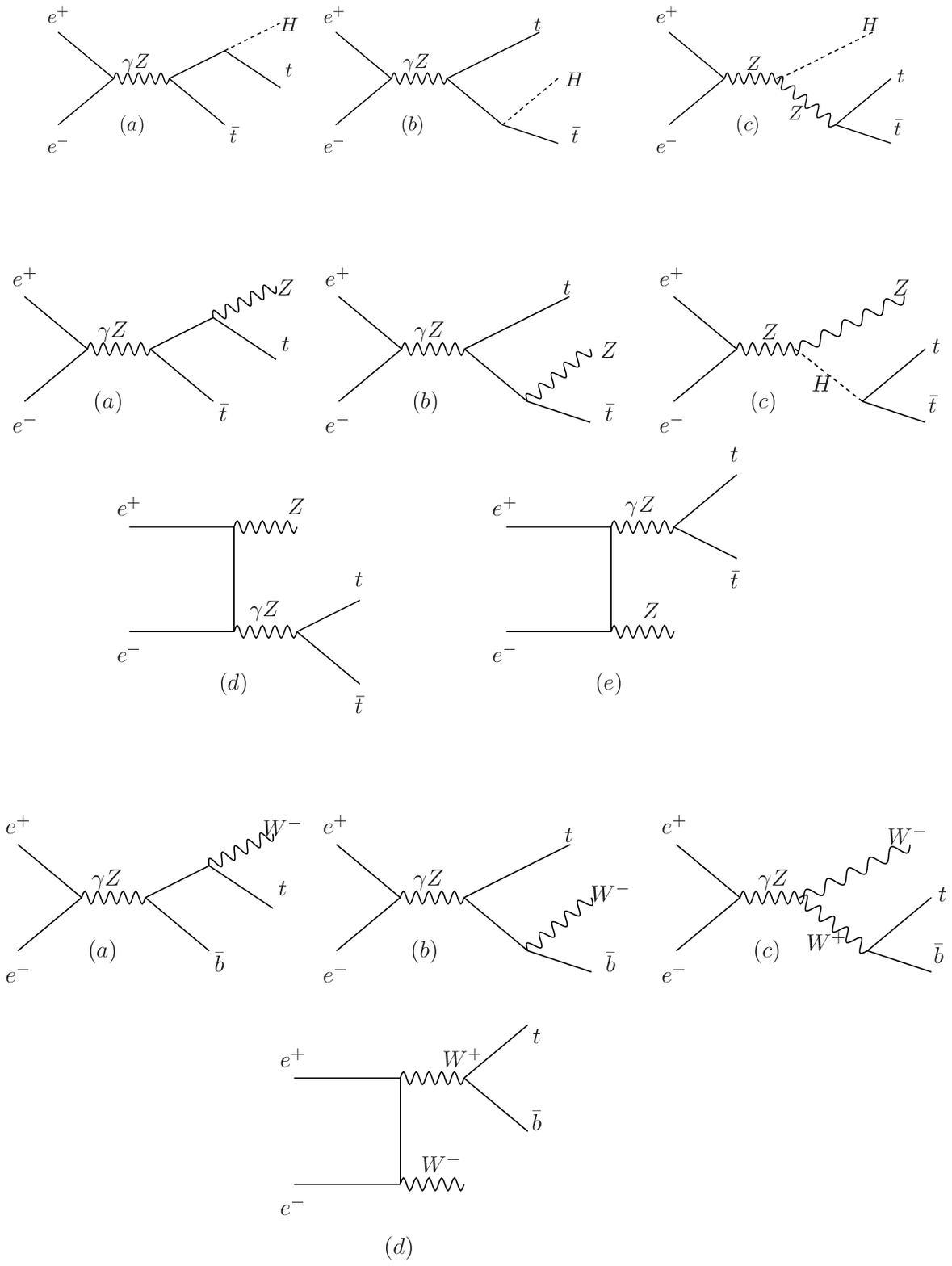, height=15.cm}
\]\\
\caption[1] {SM diagrams for $e^+e^-\to t\bar t H, t\bar t Z, t\bar b W^-$.}
\end{figure}

\clearpage

\begin{figure}[p]
\vspace{-6cm}
\[
\epsfig{file=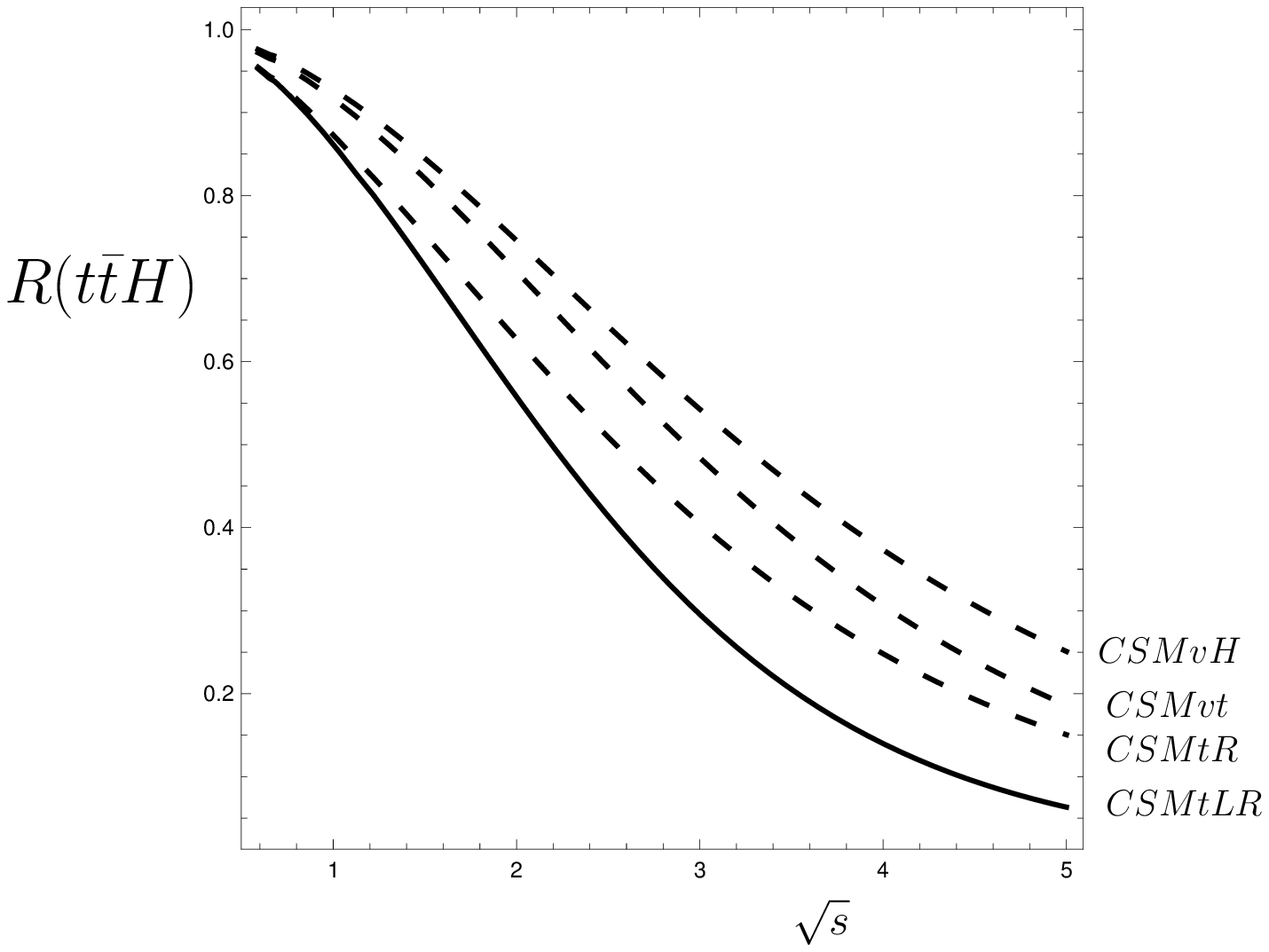, height=10.cm}
\]\\
\caption[1] {$e^+e^-\to t\bar t H$ ratios}
\end{figure}

\clearpage

\begin{figure}[p]
\[
\epsfig{file=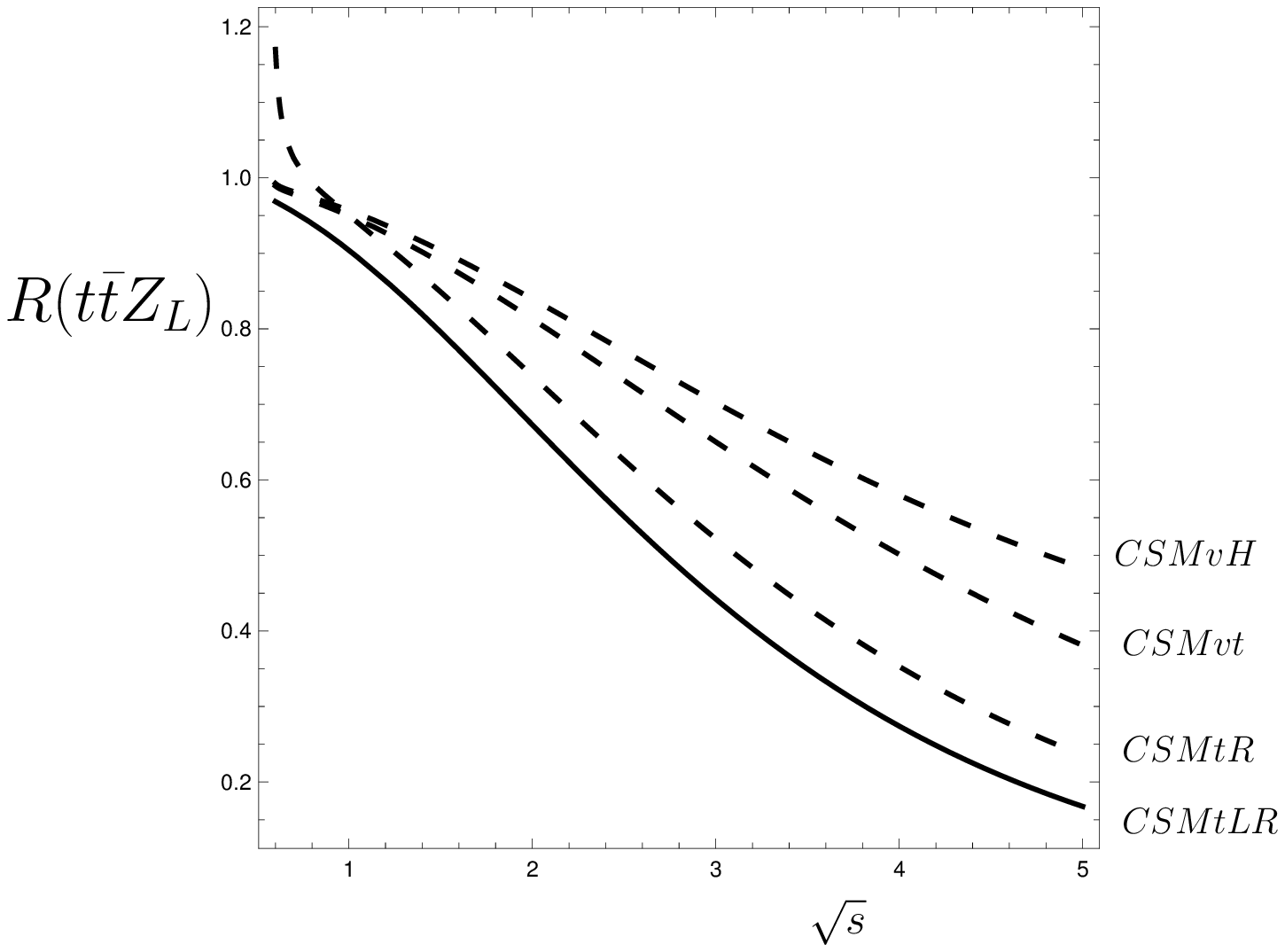, height=6.cm}\hspace{0.5cm}
\epsfig{file=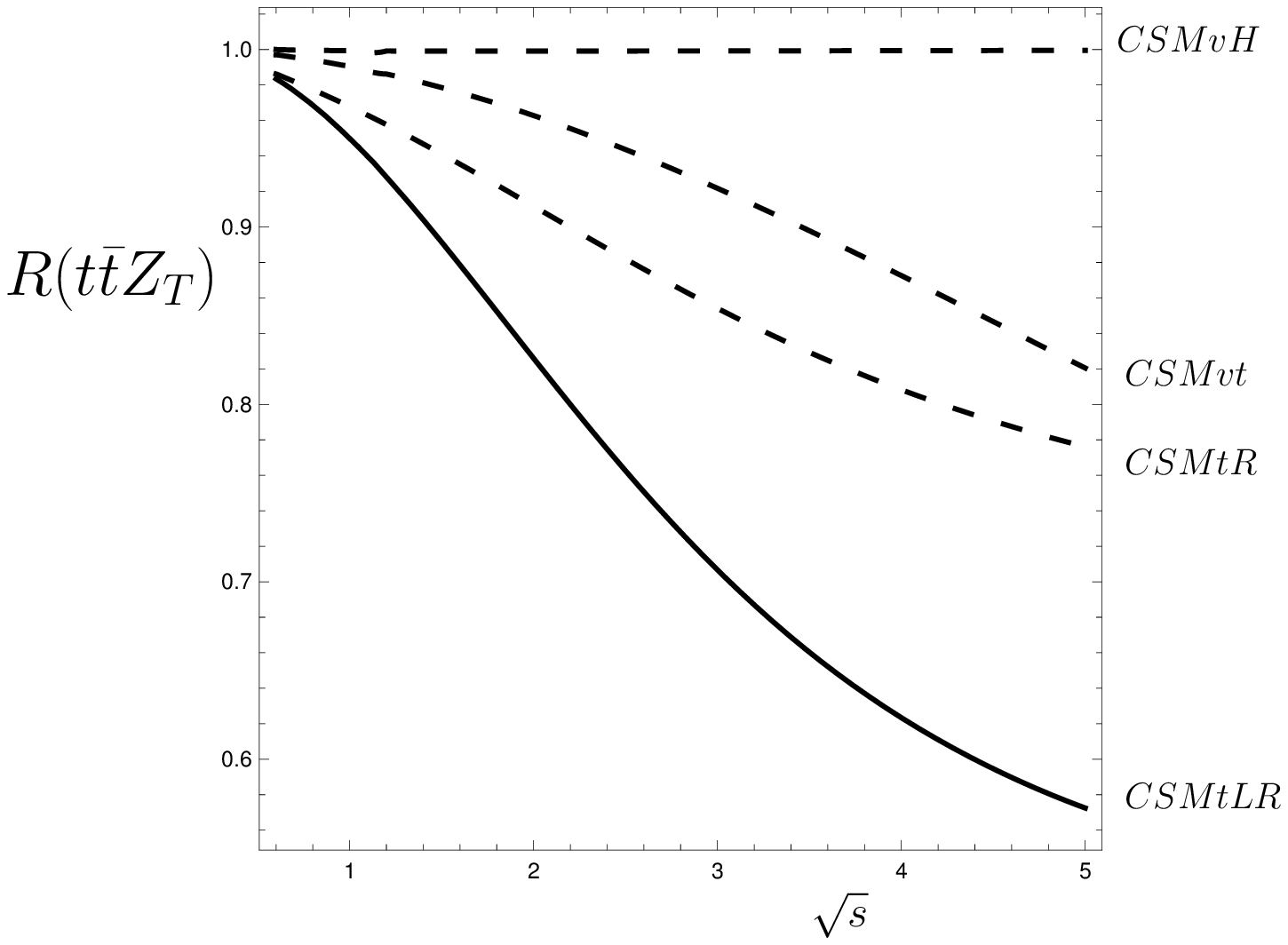,height=6.cm}
\]\\
\[
\epsfig{file=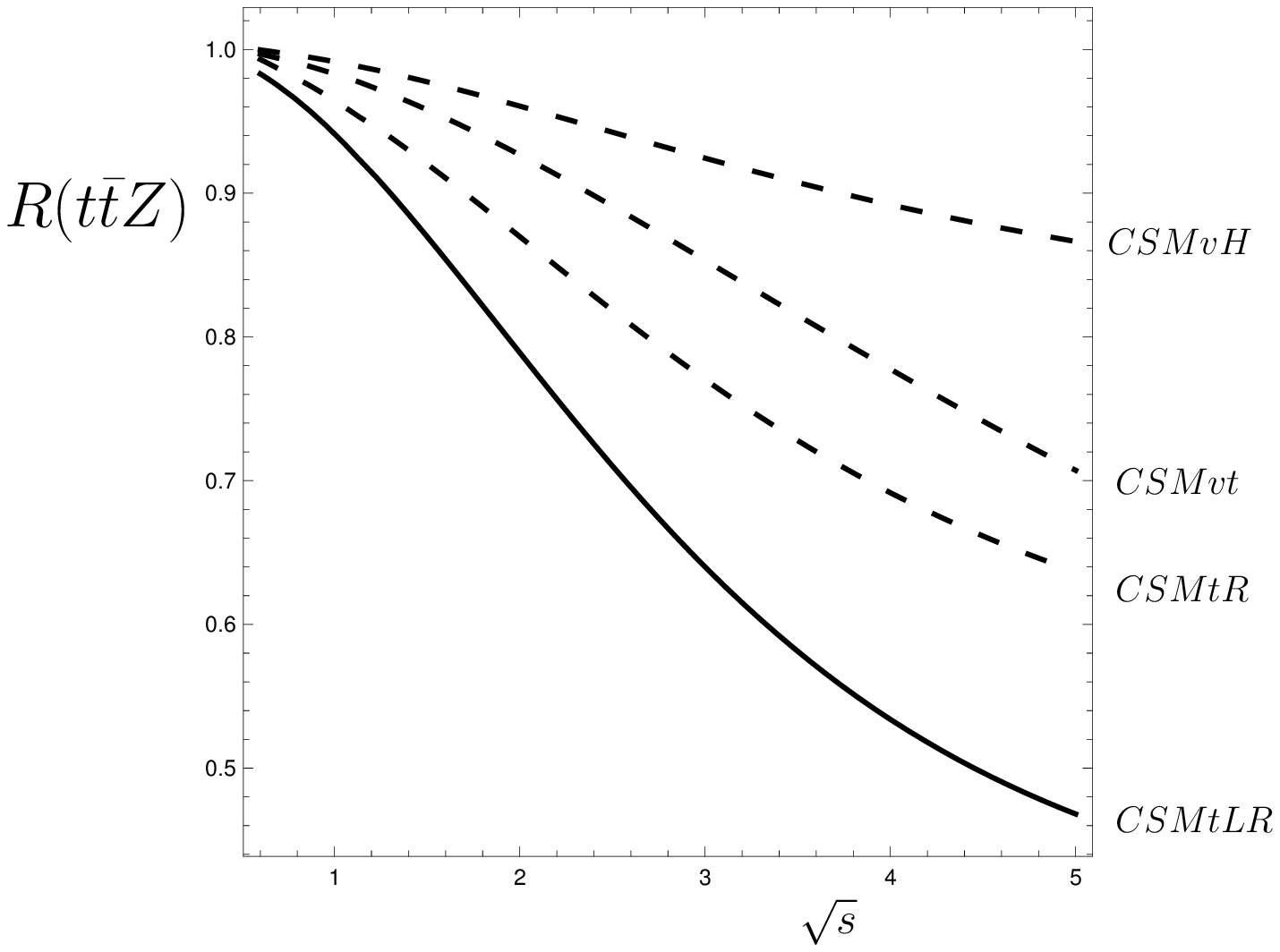, height=8.cm}
\]\\
\caption[1] {$e^+e^-\to t\bar t Z$ ratios for $Z_L$, $Z_T$ and unpolarized $Z$.}
\end{figure}

\clearpage

\begin{figure}[p]
\[
\epsfig{file=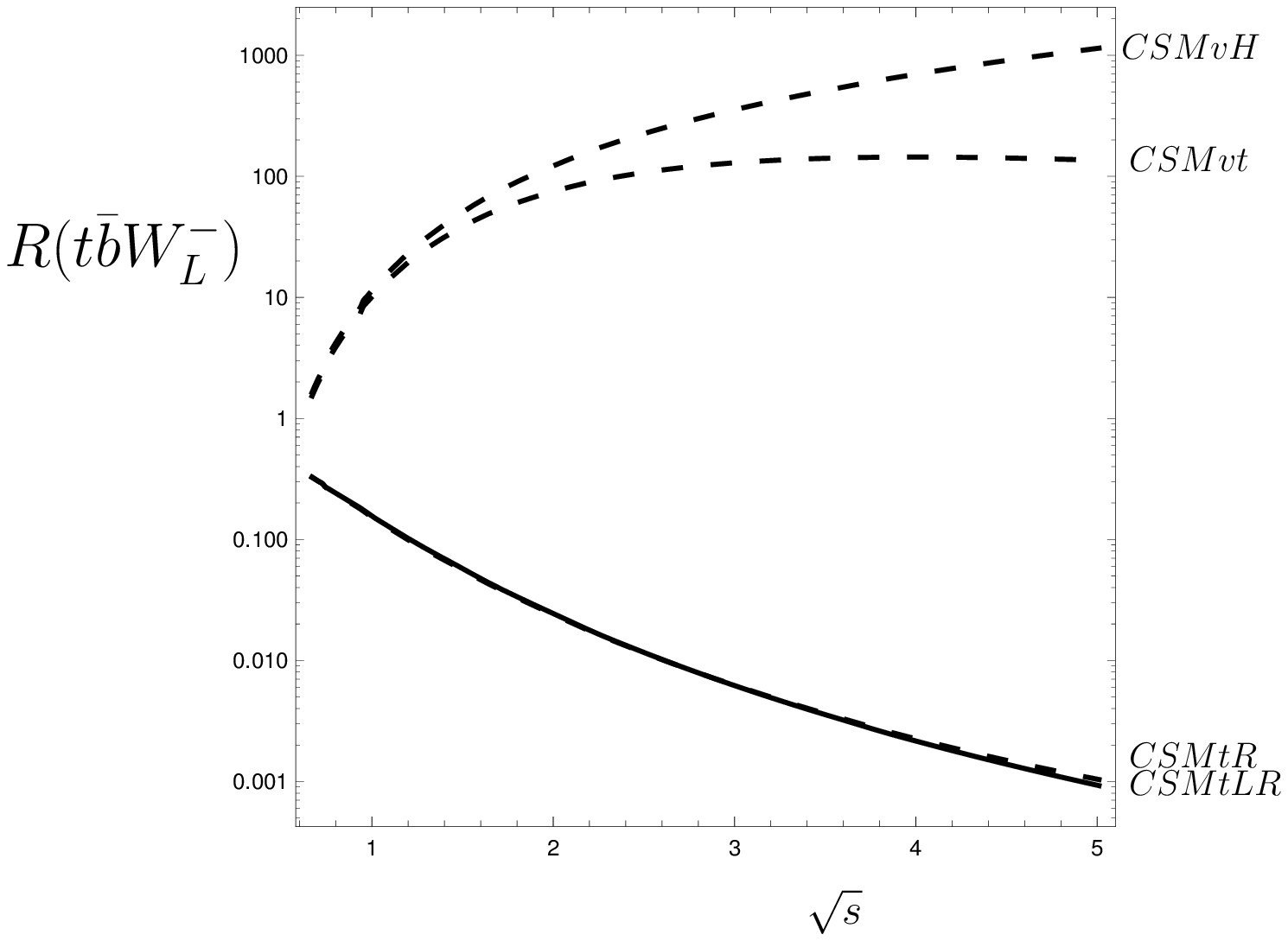, height=6.cm}\hspace{0.5cm}
\epsfig{file=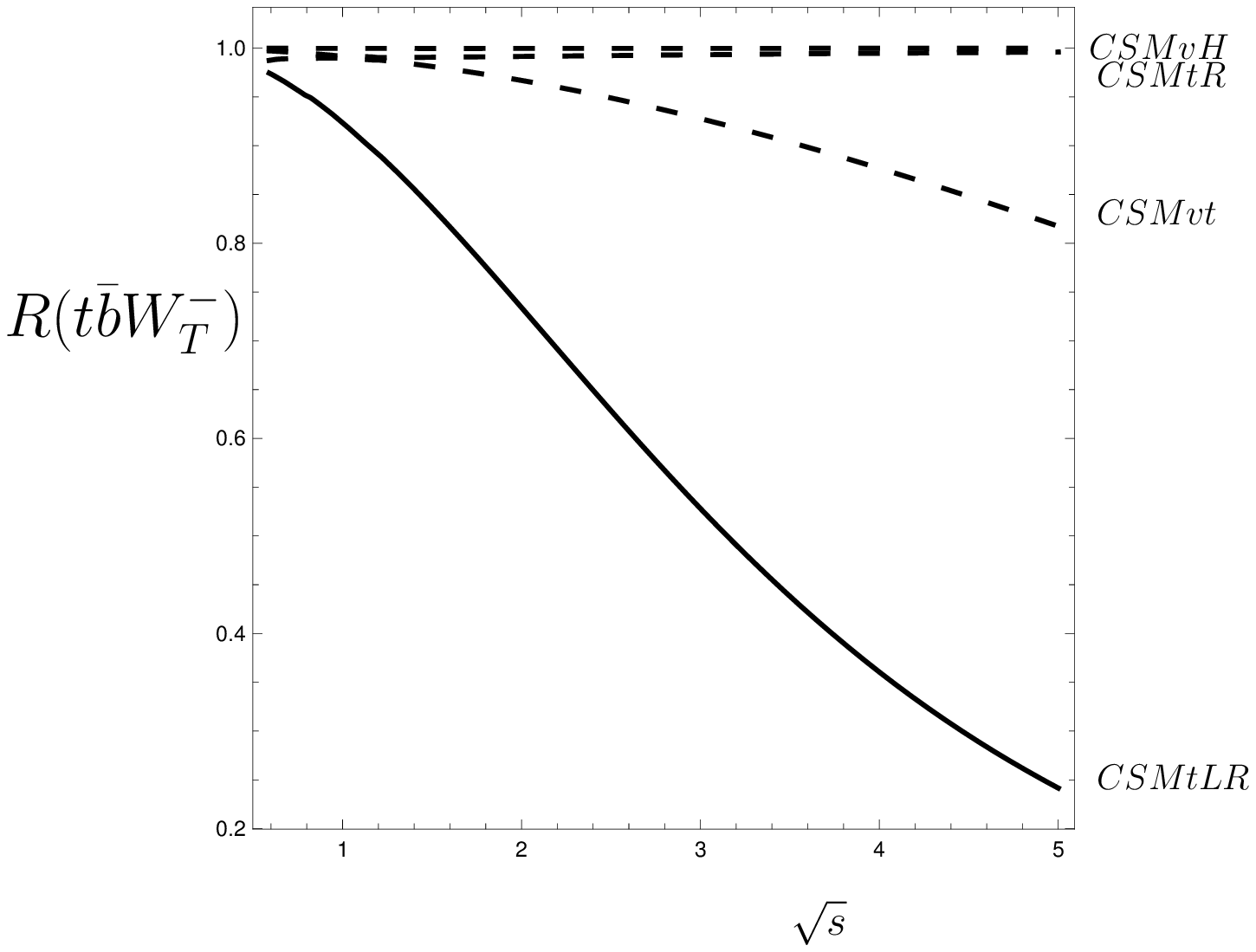,height=6.cm}
\]\\
\[
\epsfig{file=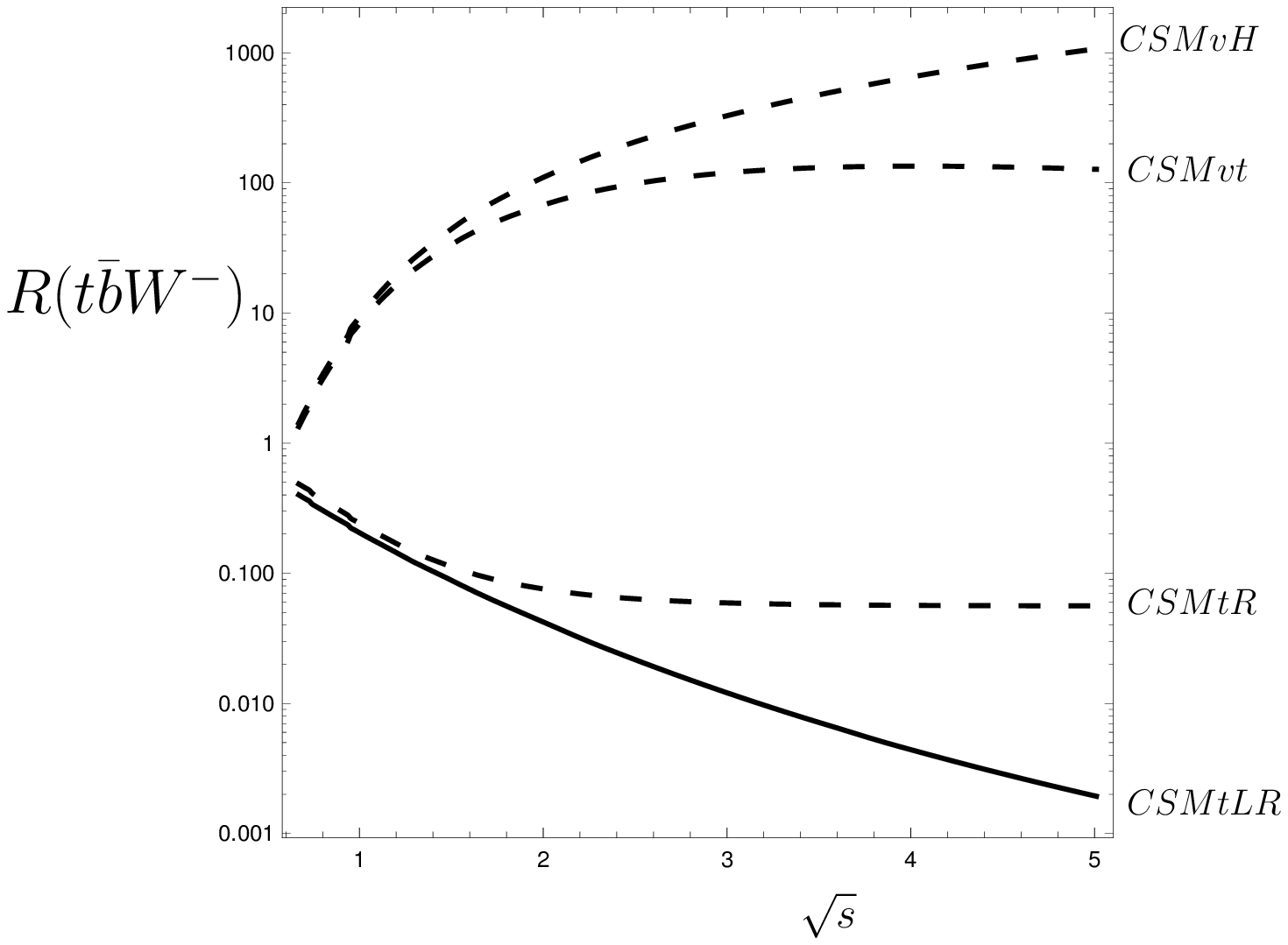, height=8.cm}
\]\\
\caption[1] {$e^+e^-\to t\bar b W^-$ ratios for $W^-_L$, $W^-_T$ and unpolarized $W^-$.}
\end{figure}

\clearpage
\end{document}